\documentclass[12pt,preprint]{aastex}
\def\t0{\theta_{\circ}}

\def\be{\begin{equation}}
\def\en{\end{equation}}

\def\rsun{R_{\sun}}

\newcommand{\msun}{$M_{\odot}$}

\newcommand{\asecdot}[2]{\mbox{#1$\stackrel {\prime \prime}{_{\bf \cdot}}$#2}}

\begin{document}

\shorttitle{An Edge-On Disk in MBM~12A}
\shortauthors{Jayawardhana et al.}

\title{Discovery of an Edge-On Disk in the MBM~12 Young Association}

\author{Ray Jayawardhana}
\affil{Department of Astronomy, University of California at Berkeley, Berkeley, CA 94720, U.S.A.}

\author{K. L. Luhman}
\affil{Harvard-Smithsonian Center for Astrophysics, 60 Garden St., Cambridge, MA 02138, U.S.A.}

\author{Paola D'Alessio}
\affil{Instituto de Astronomia UNAM, Apartado Postal 72-3 (Xangari), 58089 Morelia, Michoacan, Mexico}

\and

\author{John R. Stauffer}
\affil{SIRTF Science Center, Caltech MS 220-6, Pasadena, CA 91125, U.S.A.}

\begin{abstract}
We report the discovery of a spatially-resolved edge-on protoplanetary disk
in the $\sim$2-Myr-old MBM~12 young association. Our near-infrared images of 
LkH$\alpha$ 263C (MBM~12A~3C), obtained with the Hokupa'a adaptive optics 
system on the Gemini North telescope, clearly show two elongated reflection 
nebulosities separated by a dark lane, a morphology well-matched by scattered 
light models of an optically thick (at near-infrared wavelengths) edge-on disk.
An optical spectrum of the scattered light nebulosity obtained with the 
Keck II telescope exhibits a spectral type of M0$\pm0.5$ 
($T_{\rm eff}=3850\pm100$~K) for the central star and contains H$\alpha$ 
and forbidden emission lines, which may indicate the presence of a jet. 
The absence of a near-infrared point source implies $A_K>9.5$ toward the 
unseen central star. The disk is flared and has a radius of $\sim$150 AU 
(at a distance of 275 pc) and an inclination of 87 degrees. The aspect ratio 
of the model disk in the $J-$band is 0.72. There is possible evidence for 
dust settling to the disk midplane. LkH$\alpha$ 263C is $4\farcs115$ from 
the $0\farcs415$ binary LkH$\alpha$ 263 A and B (MBM~12A~3A and 3B), which 
is itself $15\farcs5$ from LkH$\alpha$ 262 (MBM~12A~2). Thus, LkH$\alpha$ 263C
may be the first disk to be clearly resolved around an individual star in 
a young quadruple system. The detection of a faint edge-on disk near a bright 
star demonstrates both the high angular resolution and the high sensitivity 
that can be achieved with adaptive optics imaging on large telescopes.

\end{abstract}


\section{Introduction}

Planets form out of circumstellar disks. While a vast majority of very 
young stars show indirect evidence of disks, such as excess emission in 
the infrared (IR) (e.g., Lada et al. 2000), only a handful have been imaged 
directly. Spatially-resolved images are extremely valuable for 
investigating the structure and physical properties of protoplanetary 
disks as well as their subsequent evolution into planetary systems. 
Therefore, scattered light disk images obtained by the {\it Hubble Space 
Telescope (HST)} have been the subject of intense scrutiny 
(e.g., O'Dell, Wen, \& Hu 1993; Stapelfeldt et al. 1998; Padgett et al. 
1999). Large ground-based telescopes equipped with adaptive 
optics (AO) systems are now able to achieve angular resolution and 
sensitivity comparable to that from space. Taking advantage of these new 
capabilities, we have initiated AO observations of nearby young star 
environs. Here we report the discovery of an edge-on disk in the MBM~12 
young association.

MBM~12 (=L1457; Magnani, Blitz, \& Mundy 1985) is one of the few clouds
at high Galactic latitude known to harbor young stellar objects. Until
recently, it was also considered to be the nearest molecular cloud to the
Sun, at a distance of 50-100~pc (Hobbs Blitz, \& Magnani 1986; Hearty et 
al.\ 2000). However, a recent distance estimate by Luhman (2001), 
based on an alternate analysis of the data from those studies and a new 
comparison of M dwarfs in the foreground and background of the cloud 
with local field dwarfs, places MBM~12 at $\sim$ 275 pc.
A dozen young stars have been found in projection against the cloud
and are believed to comprise a young association (MBM~12A; Luhman 2001, 
references therein).  The strengths of photospheric 
lithium absorption as well as the positions of MBM~12A stars on the H-R 
diagram suggest an age of $\sim$2 Myr for the group.
Mid-IR photometry of the original eight members reveal 
that six of them are surrounded by optically-thick accretion disks 
(Jayawardhana et al. 2001). The disk fraction found in MBM~12A --$\sim$75\%-- 
falls in the middle of the range reported for other star-forming regions 
like Taurus and Trapezium (Lada et al. 2000 and references therein), but 
is significantly higher than in $\sim$10-Myr-old nearby groups such as the 
TW Hydrae Association (Jayawardhana et al. 1999). 
Given their relative proximity, MBM~12A stars are suitable targets for 
high-resolution studies of disks and close companions. 

\section{Observations and Data Analysis}

\subsection{Adaptive Optics Imaging}

The young stars LkH$\alpha$ 262 (MBM~12A~2) and LkH$\alpha$ 263 (MBM~12A~3), 
which form a wide binary system ($15\farcs5$), were observed during the nights
of 2001 October 8 and 2000 December 25, respectively, with the University of 
Hawaii Hokupa'a AO system in conjunction with the 1024$\times$1024 array 
near-IR camera (Graves et al. 1998) on the Gemini North telescope. The 
$H$-band plate scale of QUIRC was $0\farcs01998\pm0\farcs0008$~pixel$^{-1}$
(F. Rigaut 2001, private communication), corresponding to a total field of
$20\farcs46\times20\farcs46$. Short and long exposures were obtained for 
LkH$\alpha$ 262 at $K\arcmin$ and for LkH$\alpha$ 263 at $J$, $H$, and 
$K\arcmin$. The observing and data reduction procedures will be described 
in a later study. The unsaturated point sources in the final images of 
LkH$\alpha$ 263 exhibited FWHM$=\asecdot011$ in $J$, $\asecdot0096$ in $H$, 
and $\asecdot009$ in $K'$, and are used to estimate the point spread function 
(PSF) for the disk modeling later in this work.


Gemini AO images of LkH$\alpha$ 262 and LkH$\alpha$ 263 are shown in Figure~1. 
Only one point source was found in the data for LkH$\alpha$ 262. 
LkH$\alpha$ 263 is resolved into a close binary ($0\farcs4$) with a fainter, 
extended companion ($4\arcsec$). Henceforth, we refer to the southwest and 
northeast components of the binary and the extended source as LkH$\alpha$ 
263 A, B and C (MBM~12A~3A, 3B, and 3C), in order of $K\arcmin$-band 
brightness.

Array coordinates and photometry for the A, B, and C components of LkH$\alpha$
263 were measured with the IRAF tasks IMEXAMINE and PHOT, respectively.
Aperture photometry for LkH$\alpha$ 263A and LkH$\alpha$ 263B was extracted 
with a radius of 6 pixels, which was chosen to be small enough to avoid 
significant contamination by the other component of the binary. These 
measurements were used to compute the flux ratio of the two stars in each band.
Photometric calibration was derived by combining the $J$, $H$, and $K_s$
measurements for LkH$\alpha$ 263A+B from the Two-Micron All-Sky Survey (2MASS)
with aperture photometry with a radius of 60 pixels for this system in
Gemini short exposures. This calibration was then used to measure absolute
photometry for LkH$\alpha$ 263C with an aperture radius of 40 pixels in the 
short and long exposures. The signal-to-noise of LkH$\alpha$ 263C was low in 
the short exposures, but the photometric calibrator LkH$\alpha$ 263A+B was 
not saturated in those frames, allowing accurate (if not precise) photometry. 
Meanwhile, because the conditions were not photometric and the calibration 
was derived from the short exposures, the accuracy of the measurements from 
the long exposures was compromised.  After comparing the measurements from 
the long and short exposures, we arrive at the photometry for LkH$\alpha$ 263C 
that is listed in Table~1. 

\subsection{Optical Spectroscopy}

We obtained an optical spectrum of LkH$\alpha$ 263C on 2001 November 13 with 
the Keck low-resolution imaging spectrometer (LRIS; Oke et al.\ 1995).
The long-slit mode of LRIS was used with the 400~l~mm$^{-1}$ grating 
($\lambda_{\rm blaze}=8500$~\AA) and GG495 blocking filter.
The slit width was $1\farcs0$, producing a spectral resolution of FWHM=6~\AA.
The exposure time was 600~s. After bias subtraction and flat-fielding,
the spectrum was extracted and calibrated in wavelength with arc lamp data. 
The spectrum was then corrected for the sensitivity function, which was
measured from observations of the spectrophotometric standard star Feige~110. 
The portion of the LRIS spectrum exhibiting spectral features is provided in 
Figure~2.

\section{Results}

As shown in the Gemini AO images in Figure~1, LkH$\alpha$ 263C is extended and
not associated with a point source at near-IR wavelengths. Instead, there 
are two parallel, elongated nebulosities separated by a narrow dark lane. 
The two nebulosities are about $1\farcs1$ (300 AU) in length. The southwest 
component is $\approx$20\% brighter than its northeast counterpart
in $H$-band peak brightness.
The morphology of LkH$\alpha$ 263C matches that expected for an optically thick
circumstellar disk seen nearly edge-on. The central star is hidden behind
the disk material, but its light is scattered by the upper and lower
surfaces of the disk, producing the observed nebulosities. Similar nearly
edge-on disks have been detected previously by {\em HST}, speckle, AO 
imaging around a number of young stars including HH30 (Burrows et al.\ 1996),
HK Tau B (Stapelfeldt et al.\ 1998; Koresko 1998), and HV Tau C 
(Monin \& Bouvier 2000).


The spectrum in Figure 2 is useful in constraining the physical properties of
LkH$\alpha$ 263C. Both photospheric absorption features (CaH, TiO, Na) and
a variety of forbidden emission lines ([O~I], [O~II], [N~II], [S~II]) are
detected. The continuum is well-matched by a spectral type of M0$\pm$0.5
with no reddening ($A_V<1$) and veiling that becomes detectable
blueward of 7000~\AA. Similar blue continuum excesses are commonly observed
in active T Tauri stars (e.g., Hartigan, Edwards, \& Ghandour 1995).
Comparison of our derived spectral type with the temperature scale of 
Schmidt-Kaler (1982) yields $T_{\rm eff}=3850\pm100$~K for the star. 
The lack of reddening in the spectrum and the IR colors of LkH$\alpha$ 263C 
indicates that all of the detected light is scattered, which is consistent 
with the lack of a point source in the images.
For a spectral type of M0 and an age of $<3$~Myr, the evolutionary models
of Baraffe et al.\ (1998) imply a mass of $\sim0.7$~\msun.
Meanwhile, the composite of LkH$\alpha$ 263A+B exhibits a spectral type of M3
(Luhman 2001). Because these two objects have comparable brightnesses at
IR wavelengths, they probably have similar spectral types and masses. If
we assume each has a spectral type of M3 and correct the bolometric luminosity
of the composite system (Luhman 2001) for the binarity, then they have masses 
of $\sim0.45$~\msun\ by the models of Baraffe et al. (1998). Thus, 3C is more 
massive than 3A and 3B even though it is much fainter from our vantage point. 
The H$\alpha$ and forbidden emission lines found in LkH$\alpha$ 263C
have also been observed for other edge-on systems, such as HH30 and 
HV Tau C (Kenyon et al. 1998; Magazz\'{u} \& Mart{\'\i}n 1994). 
Forbidden transitions are typical of Herbig-Haro objects, Class~I sources,
and some T Tauri stars, where the emission is believed to arise predominantly 
in jets (Kenyon et al.\ 1998, references therein).

Now that we have an estimate of the mass of the central star in LkH$\alpha$ 
263C, we can use the absence a near-IR point source to arrive at a lower 
limit for the star's line-of-sight reddening. After inserting artificial 
point sources 
of various brightnesses in the center of the dark lane in the deep $K\arcmin$ 
image, we find a detection limit of $K\arcmin\sim19$. Given the unreddened 
brightness of $K=9.5$ for an M0 star at the age and distance of MBM~12A (e.g.,
LkH$\alpha$ 262), the unseen central star must have a reddening of $A_K>9.5$. 
On the other hand, Luhman (2001) derived an extinction of $A_V\sim0$ for the 
LkH$\alpha$ 263A+B binary, just $4\arcsec$ away. Thus, the extinction toward 
the C component must be almost entirely local, i.e., due to an edge-on 
optically thick disk blocking light from the central star. 

\section{Disk Model}

We have attempted to fit a physical model to the disk around LkH$\alpha$ 263C,
following the methods described in D'Alessio et al. (2001). 
The model assumes complete mixing between dust and gas and solves for the 
vertical disk structure including the heating effects of stellar radiation 
as well as viscous dissipation in the disk itself. Fig. 3 shows the comparison
of the observations to the images of a disk model with an accretion rate of 
$5\times10^{-10}$~\msun, viscosity parameter $\alpha=0.01$, $T_*$=3850 K, 
$M_*=0.7$~\msun, $R_*=2.2$~$\rsun$, a disk radius of 150~AU, and an 
inclination angle of 87 degrees. The aspect ratio in the $J$-band model image 
is 0.72, considering the lowest contour plotted. The dust grains are assumed 
to have a powerlaw size distribution with an index of -3.5, a maximum particle
radius of 1 mm, and abundances of Pollack et al. (1994). The model disk has a 
mass of 0.0018 \msun, though this estimate is highly sensitive to the assumed 
dust opacity. The model fit is by no means unique, and while the model 
shown here fits the disk shape reasonably well, its maximum brightness is
about a magnitude higher than in the observed image. 

We have explored a wide range of parameter space for the models, and find that
the match between well-mixed models and observations is less than ideal. 
If one makes the model disk more massive or makes the grains smaller (so
that opacity in the near-IR is higher) to increase its self absorption and 
decrease the maximum brightness, the width of the dark lane grows too large.
On the other hand, if one makes the model disk less massive or the maximum 
grain size bigger, in order to make the disk less efficient at scattering 
stellar radiation, it is also less opaque and the maximum brightness becomes
too high, because a larger fraction of direct stellar radiation can be 
transported through the disk. One plausible explanation for the discrepancy
is dust settling
to the midplane of the disk. In that case, the midplane dust contributes to 
the absorption of the stellar radiation in the dark lane, but not much to the 
emission of the nebulosities, above and below the dark lane. The present 
models, which assume that dust and gas are well-mixed, cannot self-consistently
account for such an effect. Observations at millimeter wavelengths could 
provide a better mass estimate for the disk. 

\section{Discussion}

LkH$\alpha$ 263C appears to be a member of a young quadruple system in the 
MBM~12 association, along with LkH$\alpha$ 263A+B and LkH$\alpha$ 262. The 
possibility of a chance alignment of these four young stars -- particularly 
in a hierarchical fashion -- within a $20\arcsec\times20\arcsec$ region is 
extremely small considering the low stellar density of the association 
(0.002~arcmin$^{-2}$). However, because the projected separation of 
$\sim4000$~AU between LkH$\alpha$ 263 and LkH$\alpha$ 262 approaches the upper 
limit of stable binaries (see Duquennoy \& Mayor 1991), the latter component 
may become unbound from the quadruple in the future. 
Nevertheless, LkH$\alpha$ 263C is the third edge-on disk found in a multiple 
system and the first in a possible quadruple system. Components LkH$\alpha$ 
263A+B and LkH$\alpha$ 262 are also likely to harbor disks since they both 
show significant excess in the mid-infrared (Jayawardhana et al. 2001). It is 
not yet clear whether the dust in LkH$\alpha$ 263 A+B resides around one or 
both components or in a circumbinary configuration. 

While a majority of T~Tauri stars are binaries, higher order multiple 
systems appear to be relatively rare (Ghez et al. 1997 and references 
therein). The formation of multiple stellar systems is still poorly 
understood. Currently, the most promising mechanism appears to be 
fragmentation during the collapse of dense molecular cloud cores 
(Bodenheimer et al. 2000). High spatial resolution observations of 
pre-main-sequence multiple systems, like those presented here, help 
determine the stellar and disk configurations of young objects, and can 
lead to better constraints on models of multiple star formation and early 
evolution. The fact that the disk of LkH$\alpha$ 263C is clearly neither 
coplanar nor parallel with its orbital plane is a particularly useful test 
for such models.

\acknowledgements
We with to thank Kathy Roth, Francois Rigaut, Mark Chun, Olivier Guyon, and 
Dan Potter for their assistance in obtaining the observations at Gemini 
Observatory. We also thank Davy Kirkpatrick and the staff of Keck Observatory 
for their assistance in the Keck observations. We are grateful to Javier 
Ballesteros for his help with the modeling.


\clearpage
\begin{figure}
\plotone{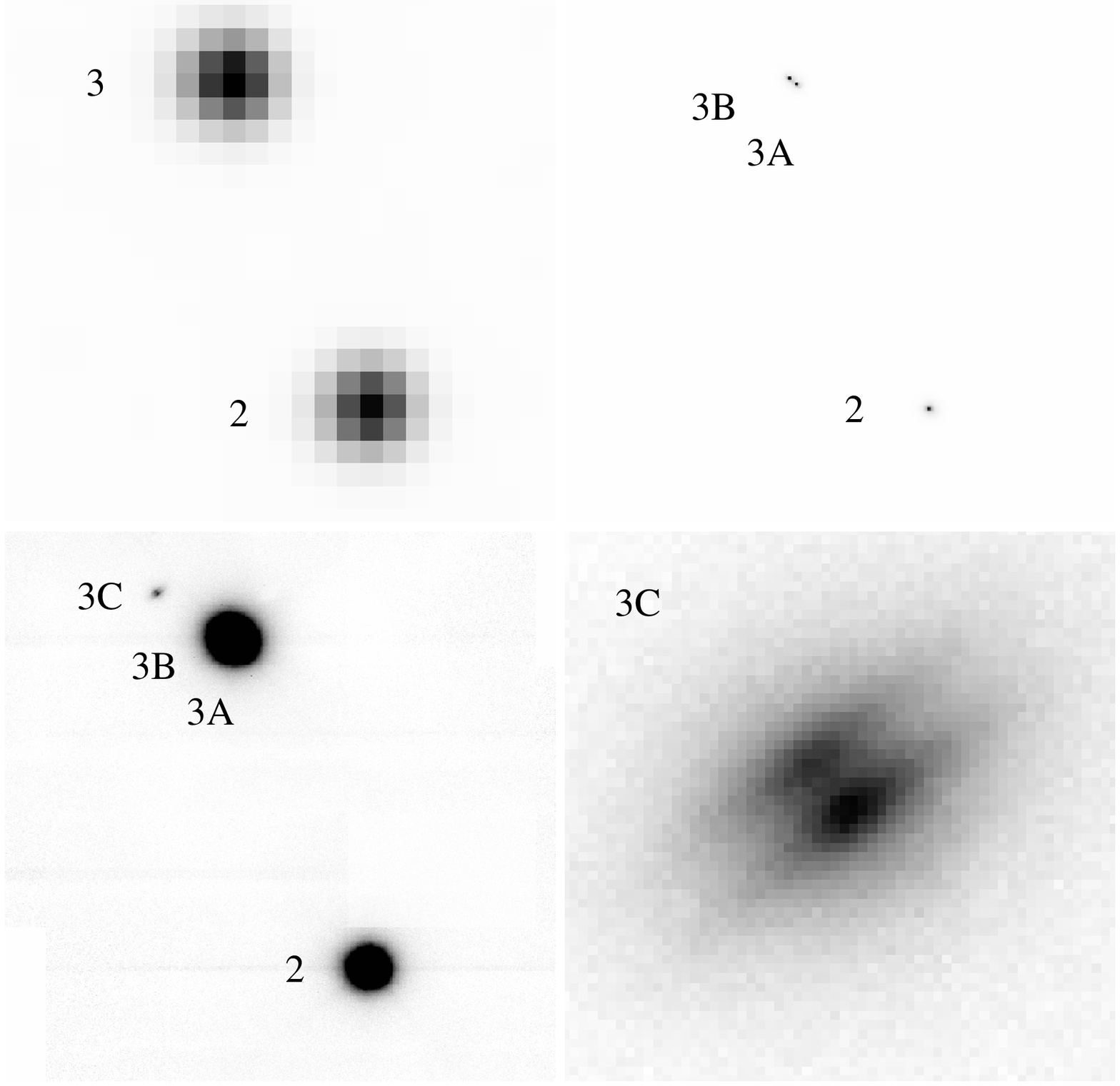}
\caption{
{\em Upper left}: A 24$\arcsec\times24\arcsec$ image in $K_s$ from 2MASS
shows the young wide binary LkH$\alpha$ 262 (MBM12A~2) and LkH$\alpha$ 263 
(MBM12A~3) separated by $15\farcs5$. 
{\em Upper right}: Short $K\arcmin$ exposures with Gemini AO for the same 
field. LkH$\alpha$ 263 is resolved into a close binary ($0\farcs42$).
{\em Lower left}: Deep $K\arcmin$ exposures reveal a faint (H=16), 
nebulous object $4\arcsec$ north-east of LkH$\alpha$ 263. 
{\em Lower right}: The $1\farcs2\times1\farcs2$ region surrounding 
LkH$\alpha$ 263C in a deep $H$-band image clearly shows an edge-on disk 
morphology.}
\end{figure}

\clearpage
\begin{figure}
\plotone{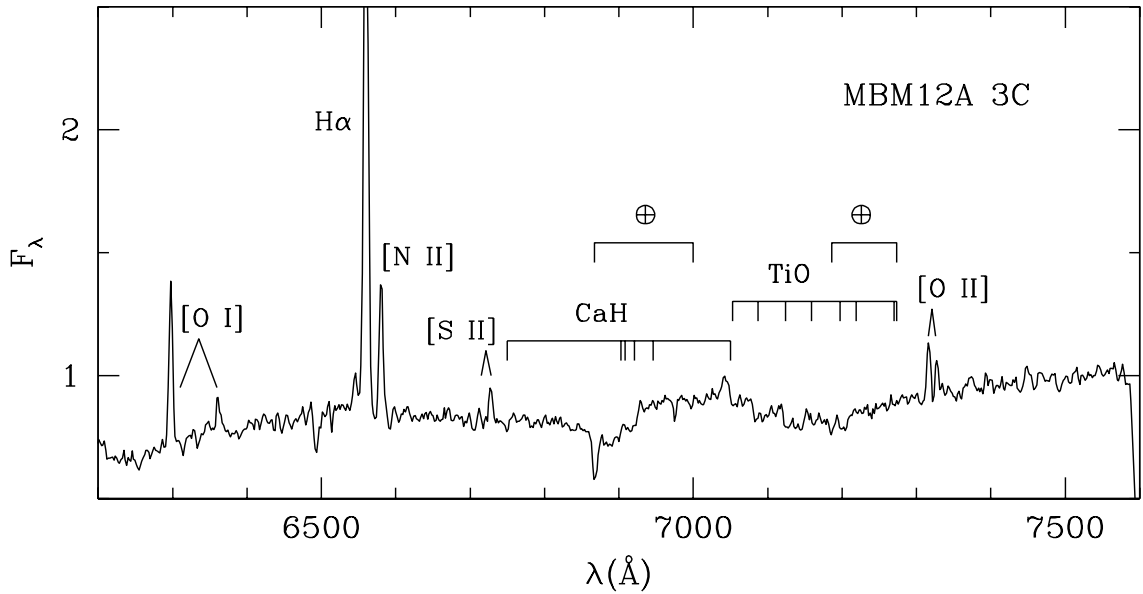}
\caption{Optical spectrum of the companion edge-on disk, LkH$\alpha$ 263C. 
These data indicate a spectral type of M0$\pm$0.5 with blue continuum veiling 
at $<7000$~\AA. The forbidden emission lines in this object are commonly 
observed toward Herbig-Haro objects, Class I sources, and some 
T Tauri stars and may arise from a jet.
For 7600-9000~\AA, the spectrum of this object is featureless 
except for Na absorption near 8200~\AA. The spectrum is normalized at 7500~\AA.
}
\end{figure}

\clearpage
\begin{figure}
\plotone{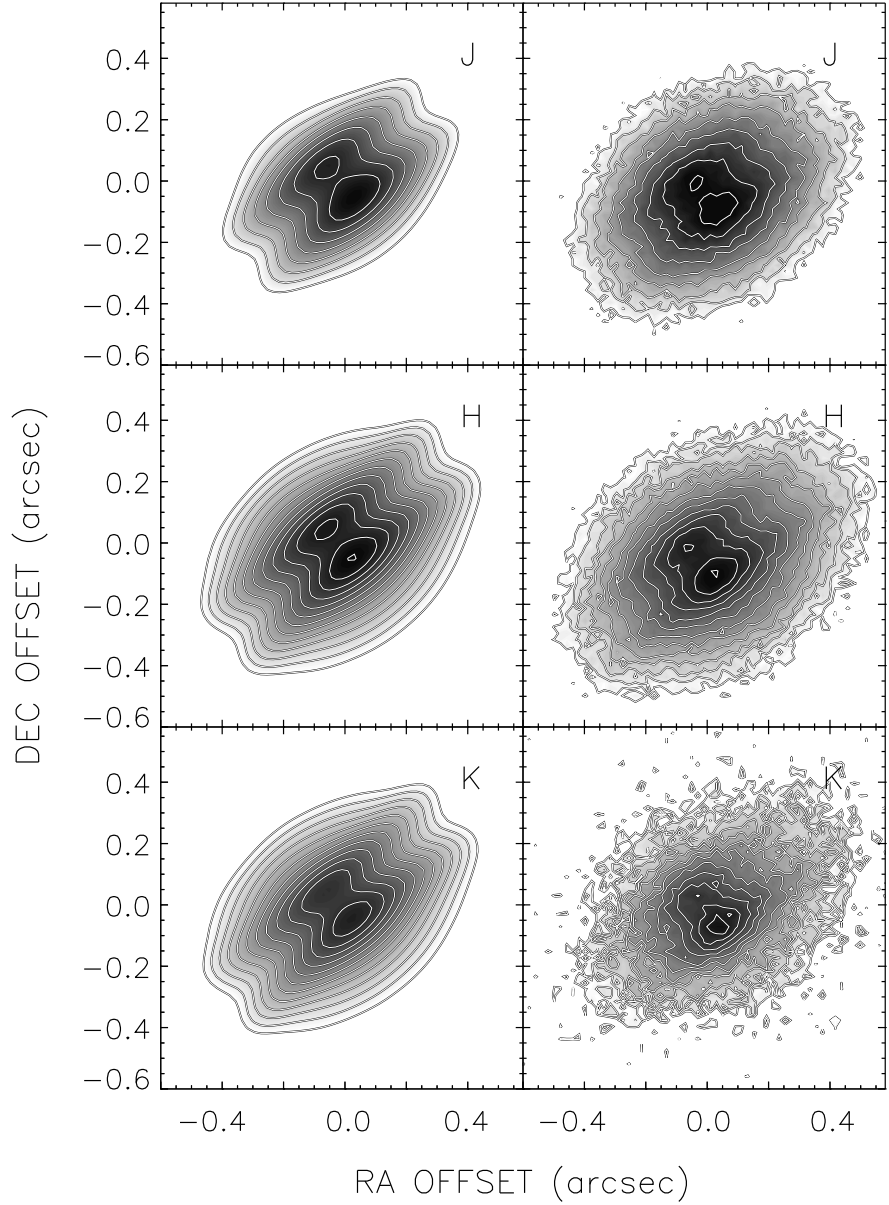}
\caption{Comparison of the model ({\em left}) and observed ({\em right})
images of the LkH$\alpha$ 263C disk in $J$, $H$ and $K\arcmin$ from top to 
bottom, respectively. Contour levels in both images are in steps of 0.2 
mag/arcsec$^2$. See text for the model parameters and discussion.}
\end{figure}
\clearpage

\begin{deluxetable}{llllll}
\tablewidth{0pt}
\tablecaption{Astrometry and Photometry of  LkH$\alpha$ 263 (MBM12A~3)}
\tablehead{
\colhead{Component} &
\colhead{Separation\tablenotemark{a}} &
\colhead{PA\tablenotemark{a}} &
\colhead{$J$\tablenotemark{b}} &
\colhead{$H$\tablenotemark{b}} &
\colhead{$K_s$\tablenotemark{b}} \\
\colhead{} &
\colhead{(arcsec)} &
\colhead{(deg)} &
\colhead{(mag)} &
\colhead{(mag)} &
\colhead{(mag)}
}
\startdata
A &  \nodata     &   \nodata          & 11.52 & 10.64 & 10.21 \\
B & 0.415$\pm$0.004 & 51.9$\pm$0.1    & 11.25 & 10.51 & 10.34 \\
C & 4.115$\pm$0.02  & 58.3$\pm$0.2    & 16.5$\pm$0.15 & 16.0$\pm$0.15 & 16.1$\pm$0.3 \\
\enddata
\tablenotetext{a}{Positions of B and C are measured with respect to A.}
\tablenotetext{b}{Photometry for A and B are computed by combining the 
photometry for A+B from 2MASS with the relative brightnesses of A and B in the
AO images.}
\end{deluxetable}

\end{document}